# ProbMetab: an R package for Bayesian probabilistic annotation of LC-MS based metabolomics

Ricardo R. Silva[1], Fabien Jourdan [2,3], Diego M. Salvanha [1,4],
Fabien Letisse [3], Emilien L. Jamin [2,3], Simone Guidetti-Gonzalez [5],
Carlos A. Labate [5,6] and Ricardo Z.N. Vêncio [1*]

[1]LabPIB, Department of Computing and Mathematics FFCLRP-USP, University of Sao Paulo, Ribeirao Preto, Brazil

[2]INRA UMR1331, Toxalim, Research Centre in Food Toxicology, Toulouse, France

[3]Université de Toulouse, INSA, UPS, INP; LISBP, Toulouse, France

[4]Institute for Systems Biology, Seattle, Washington, USA

[5]Department of Genetics ESALQ-USP, University of Sao Paulo, Piracicaba, Brazil

[6]Laboratorio Nacional de Ciencia e Tecnologia do Bioetanol CTBE, Campinas, Brazil

October 27, 2013

**Abstract**

## 1  Summary:

We present ProbMetab, an R package which promotes substantial improvement in automatic probabilistic LC-MS based metabolome annotation. The inference engine core is based on a Bayesian model implemented to: (i) allow diverse source of experimental data and metadata to be systematically incorporated into the model with alternative ways to calculate the likelihood function and; (ii) allow sensitive selection of biologically meaningful biochemical reactions databases as Dirichlet-categorical prior distribution. Additionally, to ensure result interpretation by system biologists, we display the annotation in a network where observed mass peaks are connected if their candidate metabolites are substrate/product of known biochemical reactions. This graph can be overlaid with other graph-based analysis, such as partial correlation networks, in a visualization scheme exported to Cytoscape, with web and stand alone versions.

## 2  Availability and Implementation:

ProbMetab was implemented in a modular fashion to fit together with established upstream (xcms, CAMERA, AStream, mzMatch.R, etc) and downstream R package tools (GeneNet, RCytoscape, DiffCorr, etc). ProbMetab, along with extensive documentation and case studies, is freely available under GNU license at: `http://labpib.fmrp.usp.br/methods/probmetab/`.

## 3  Supplementary information:

Supplementary data (supporting tables and two detailed case studies) are available at Bioinformatics online and at ProbMetab website.

---

[*]to whom correspondence should be addressed

## 4 Contact:

rvencio@usp.br

## 5 Introduction

Metabolomics is an emerging field of study in post-genomics, which aims at comprehensive analysis of small organic molecules in biological systems. Techniques of mass spectrometry coupled to liquid chromatography (LC-MS, Liquid Chromatography-Mass Spectrometry) stand out as dominant methods in metabolomics experiments.

The workflow in conventional mass spectrometry analysis generates a set of characteristic mass peaks associated with chromatographic retention times and requires a large labor investment for manual inspection. Also, recent advances in ultra high resolution mass spectrometers that enabled experimental accuracies of less than 1 ppm (part per million of one unit of mass) are still insufficient to distinguish among candidate compounds in high mass windows (Kind and Fiehn, 2006).

Although computational strategies have been used to filter and annotate mass peaks in LC-MS experiments (Dunn *et al.*, 2012), these methods do not include the addition of external information into a mathematical model in a principled way. Recently, (Rogers *et al.*, 2009) put forward a proof-of-concept in which information incorporated to a probabilistic model provides better annotation (Breitling *et al.*, 2013). Their Bayesian model introduces the elegant idea of using a set of known chemical reactions among candidate compounds to improve annotation since certain combinations, detected together, would make more biochemical sense than others.

The state-of-the-art in probabilistic annotation established by (Rogers *et al.*, 2009) did not include: an integrative computational implementation; a practical connection to public databases biological such as KEGG or MetaCyc (Altman *et al.*, 2013); nor a network-based output visualization schema. Therefore, our contribution is to fulfill these specific needs allowing easy access to this powerfull statistical model for all metabolomics bioinformatics community (Fig. 1).

## 6 Methods

Aiming clarity, we refer the interest reader to the original work (Rogers *et al.*, 2009) and its subsequent extention (Rogers *et al.*, 2010) for mathematical details. Here we briefly recall the essential ingredients of their model.

The surjective mapping between mass peaks onto compounds $m \mapsto c$; $m \in \{1, ..., M\}$, $c \in \{1, ..., C\}$ and $M \leq C$; is represented by a column vector $\vec{z}_m \in \{1, 0\}^C$ such that $|\vec{z}_m| = 1$. The joint distribution of all mappings given the observed mass spectra is: $f(\vec{z}_1, ..., \vec{z}_M | x_1, ..., x_M, I_1(t), ..., I_M(t), \theta_{M+1}, ..., \theta_N)$, abreviated by: $f(Z|\vec{x}, \vec{\theta})$. In this notation, $Z$ concatenates all column vectors in a matrix; $x_m$ is the $m$-th mass peak's actual observed mass and $I_m(t)$ its measured intensity (possible a time-series and/or repeated experiments); and $\vec{\theta}$ encodes all other $N$ necessary parameters such as $\vec{I}(t)$, spectrometer precision, known compounds in the biological database chosen, adjacency matrix of bioreaction graphs, hyperparameters, and so on. Gibbs sampling is performed as usual to access such complicated joint $f(\cdot)$ making use of its marginals: $f(\vec{z}_m | Z^{(-m)}, \vec{x}, \vec{\theta})$ where $Z^{(-m)}$ is $Z$ lacking the $m$−th column relative to the mass peak assignment $\vec{z}_m$. These probably distributions, by Bayes rule, are proportional to: $L(x_m | Z, \vec{x}^{(-m)}, \vec{\theta}) \cdot p(\vec{z}_m | Z^{(-m)}, \vec{x}^{(-m)}, \vec{\theta})$. $L(\cdot)$ and $p(\cdot)$ represent the likelihood and *prior* functions, respectively, and $\vec{x}^{(-m)}$ means $\vec{x}$ lacking $x_m$.

As pioneered by (Rogers *et al.*, 2009), we still use a straightforward Dirichlet-categorical *prior* $p(\cdot)$ on the chemical reaction graph to enhance the odds of more connected compounds over less connected ones. However, we integrate this model to real biochemical networks/pathways databases such as KEGG or BioCyc (Altman *et al.*, 2013) in a sensible way in order to retain biologically meaningful *priors*. Although this integration is made by web-services provided by public databases, ProbMetab allows in-house (possibly curated or specific) pathway databases to be used to inform the *prior* component of this Bayesian model.

In a subsequent review, (Rogers *et al.*, 2010) briefly suggested how their previous method could be extended to incorporate additional experimental information and metadata since the original formulation considered only a measurement noise model. We followed these leads expanding the likelihood function $L(\cdot)$ in multiplicative independent terms to account for additional orthogonal information sources:

$$L(x_m|Z, \vec{x}^{(-m)}, \vec{\theta}) \propto L_N(x_m|Z, \vec{\theta}_N) \cdot L_{rt}(x_m|Z, \vec{\theta}_{rt}) \cdot L_{iso}(x_m|Z, \vec{\theta}_{iso}) \qquad (1)$$

where $L_N(\cdot)$ is the measurement noise model; $L_{rt}(\cdot)$ is the retention time error model and; $L_{iso}(\cdot)$ is the isotope profile error model. $\vec{\theta}_\star$'s are their respective appropriate parameters subset of $\vec{\theta}$.

We implemented computationally this improved likelihood function in ProbMetab with modularity in mind. This is important because depending on particular dataset, equipment setup, experimental conditions and so on, a given multiplicative term in $L(\cdot)$ can be dropped, or a new one included.

For example, there are isotope profile error models proposed for Orbitrap (Thermo Fisher) mass spectrometry platform that can take advantage of known natural relative abundance of carbons $^{12}$C and $^{13}$C. Taking into account measured peak intensities would improve the odds of a given compound over others if it fulfills an expected ratio (Weber *et al.*, 2011).

Additionally, one may include a retention time error model to take advantage of known differences in liquid chromatography elution time of chemically distinct compounds with similar mass. Models to predict retention time exist (Creek *et al.*, 2011) although still error prone. The decision to include or drop this term, $L_{rt}(\cdot)$, is up to the practitioner and ProbMetab is equally prepared to perform probabilistic annotations with both options.

The likelihood term $L_N(\cdot)$ is mandatory and was the sole term in (Rogers *et al.*, 2009) original model's $L(\cdot)$. This measurement noise model was introduced as a gaussian density with a fixed variance $1/\gamma$ given by the mass spectrometer's precision $\gamma$. Therefore, the likelihood $L_N(\cdot)$ was proportional to $\exp(-\frac{1}{2}\gamma(x_m/y_c - 1)^2)$ and the parameters $\vec{\theta}_N = (\vec{y}, \gamma)$, where $\vec{y} = (y_1, ..., y_C)$ are the exact masses of compounds attributable to $m$, and the precision $\gamma$.

Experience shows that relying on nominal precision figures from vendors is not optimal since it can vary with several factors in an actual experimental setup and site (Böcker *et al.*, 2009). In order to overcome this limitation, we replaced $L_N(\cdot)$ following the ideas put forward by (Böcker *et al.*, 2009) when modeling isotope and noise error simultaneously but, here, considering only the noise part:

$$L_N(x_m|Z, \vec{y}, w) \propto 1_\omega \left(1 - \Phi\left(\frac{1}{\sigma_\omega}(|x_m/y_c - 1| - \mu_\omega)\right)\right) \qquad (2)$$

where $1_\omega$ is the indicator function for the set $\omega$ composed by all $j$ candidate compounds in the database for which $|x_m - y_j| \leq w$; $\mu_\omega$ and $\sigma_\omega$ are the average and standard deviation of all masses in $\vec{y}$ listed in $\omega$; $\Phi$ is the normal cumulative distribution function; and $w$ is a researcher-defined mass difference limit.

The set $\omega$, defined by the threshold $w$, is an improvement that inhibits the unwanted effect of letting highly connected compounds (hubs in the metabolic networks) overinfluence the *posterior* when using the originally proposed Dirichlet-categorical *prior*. It captures the practitioners' common understanding that, in real experimental setups, there are mass peaks that would never be confused beyond certain mass limit. The indicator function rules out unwanted compounds turning their $L_N(\cdot)$ to zero. Moreover, considering variance over $\omega$ instead of a static precision value $\gamma$, the model becomes data-centric instead of relying on equipment (ideal) specifications. These improvements come from taking (Rogers *et al.*, 2009) and (Böcker *et al.*, 2009) ideas together.

The product of a probabilistic model is a list of ranked assignments in which the probability values create a gradient among candidates. If a mass peak $m$ has only one feasible candidate $c$ (e.g. $|\omega| = 1$), there are no ties to be broken and $z_{cm} = 1$ with probability one. On the other hand, if no external information added is able to untie assignments from $m$ to $c_1, ..., c_{|\omega|}$, the probability values are all equal to the uninformative $1/|\omega|$. Also, even if there is a probability gradient among candidate compounds

for a given mass peak, it is conceivable that the true assignment could not be the $1^{st}$ ranked but rather be among the top ranked compounds. Due to several reasons, ranging from modeling limitations to measurement noise, the rankings could be imprecise and a probabilistic annotation, instead of a simple single one-to-one assignment, would be best considered as a whole.

# 7 Results and Conclusion

The platform chosen for implementation of these ideas was the well-known and established R programming environment (R Development Core Team, 2010), which incorporates a wide range of analyzes including successful tools that perform preprocessing of spectral data required for metabolite annotation (Fig. 1) (Smith *et al.*, 2006; Kuhl *et al.*, 2011).

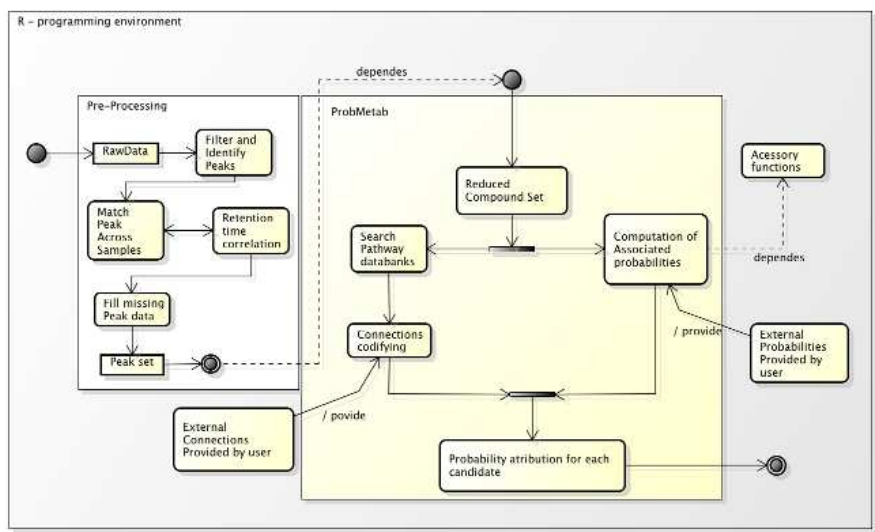

Figure 1: Schematic representation of ProbMetab package scope in spectral data analysis, showing the integration in a programming environment that provides different solutions for analysis and information integration associated with the model.

The main product of a probabilistic annotation is a list of compound candidates ranked by their probabilities (Fig. 2). In order to easily navigate over ProbMetab's results, we display tabular and dynamic network outputs along with supporting information, which assists practitioners to ultimately decide on most parsimonious annotations instead of forcing them to simplistically rely on the top probability assignment. All mass peaks are viewed as graph's nodes. Edges between two nodes are drawn if any candidate compound assigned to the outgoing node can be metabolized to any candidate compound assigned to the incoming node by means of a known biochemical reaction (Fig. 3). ProbMetab is capable of producing reaction graphs and export them as standard Cytoscape input files or broadcasting the necessary graph data and attributes (color, shapes, etc) directly to Cytoscape Desktop (Shannon *et al.*, 2013). This information can be easily overlaid with other widely used systems biology strategies such as correlation or partial correlation networks. If a mass spectra time-series or biological replicates are available, ProbMetab uses third party packages integrated downstream to export correlation or partial correlation graphs, along with their intersection/difference with the reaction graph.

Alternatively, a biologist can visualize ProbMetab's results in a simplified searchable web interface. Our package has a function that is responsible to consume an online web-service, which checks and renders the broadcast results as a web-page. The visualization approach was developed taking advantage of the cytoscape.js library (Lopes *et al.*, 2010) and its dependencies and can be easily integrated or embedded into any html5 web application.

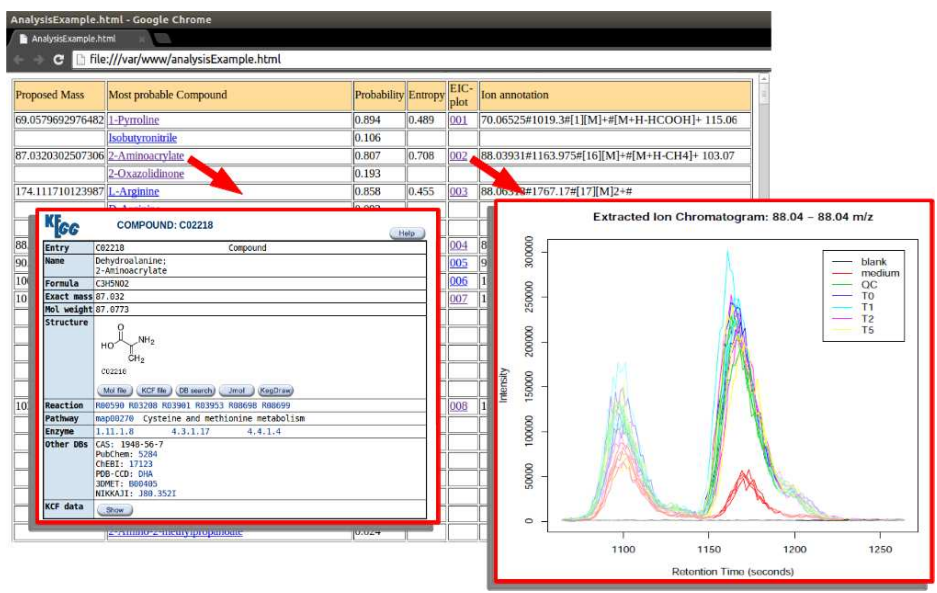

Figure 2: Screenshot of ProbMetab's tabular output showing how the probability ranking is associated to spectra and database weblinks.

ProbMetab's documentation brings two detailed case studies in which all its features are explored. Moreover, to highlight integration with downstream and upstream third party R packages, data analysis examples mentioned are carried out from raw data, following through preprocessing until it reaches ProbMetab's specific point of action. We used publicly available data from *Trypanosoma brucei*, causative agent of sleeping sickness, and an original dataset from *Sacarum oficinarum* (sugarcane), an important biofuel source, to illustrate several points in typical metabolomics analysis sections.

The *T brucei* dataset, obtained from the mzMatch.R project website, was chosen because it presents a set of metabolites identified with the aid of internal control standard compounds, being specially suited for performance evaluation. With this validation dataset we compare the MetSamp (http://www.dcs.gla.ac.uk/inference/metsamp/) implementation from (Rogers *et al.*, 2009) with ProbMetab's implementation and show that, the efficient R/c++ integrated function (Eddelbuettel and François, 2011) had a 3 fold time improvement over the MATLAB implementation. For both implementations the higher probability candidate was the true identity in up to 60% of the metabolites. However, instead of reporting only the higher probability candidate identity as proposed by (Rogers *et al.*, 2009) we show that exporting the complete ranking in summarized visualizations, up to 90% of metabolites identities are among the top-three higher probabilities. The full or filtered ranking allow the experimenter to associate the candidates with additional information present in this outputs and attribute the correct identity.

The sugarcane dataset was chosen to exemplify differential expression of annotated metabolites in contrasting environmental perturbation. We successfully recovered changes in a known stress response pathway (flavone and flavonol biosynthesis), showing the importance of a network-centric visualization for metabolite annotation in order to track metabolism changes. The benchmark dataset confirms, as preconceived by (Rogers *et al.*, 2010), that a probabilistic model using orthogonal data and metadata yields better automatic mass peak annotation. The perturbation dataset shows that probabilistic annotation can produce otherwise impossible interpretation for differential network connectivity.

We implemented a method to annotate compounds in a computational framework that allows the introduction of prior knowledge and additional spectral information. With the R package ProbMetab we provide ways to summarize the results of series of analysis needed to extract information from complex high dimensional mass spectrometry data, and help the experimenter to track metabolism changes in the process of interest.

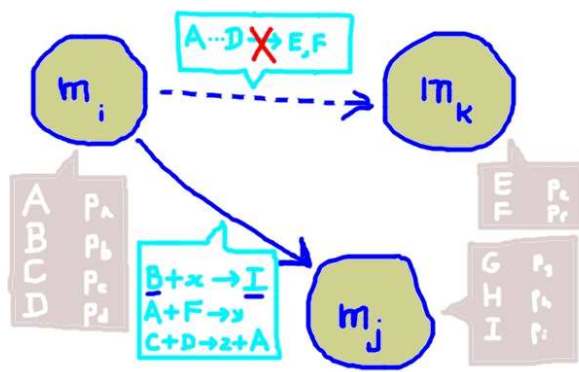

Figure 3: Conceptual representation of the network visualization for a hypothetical metabolome probabilistic annotation. Nodes are mass peaks and edges are drawn only if there is at least one biochemical reaction transforming an annotated compound into some annotated compound. Here, $m_i$, $m_j$ and $m_k$ are observed mass peaks. $A$ to $D$ are candidate compounds to annotate $m_i$ with probabilities $p_a \leq ... \leq p_d$. Analogously, $E$ to $I$ are assigned with probabilities $p_e$ to $p_i$. Since there is a known biochemical reaction metabolizing $m_i$'s $2^{nd}$ best candidate $B$ to $I$, an edge is drawn from $m_i$ to $m_j$. There are no known reactions from any $m_i$'s candidates to $m_k$'s candidates thus no edge is drawn.

# Acknowledgement

We thank Bruno M Della Vecchia (USP), Dr Oscar Yanes's team (Rovira i Virgili University) and Paul Shannon (Fred Hutchinson Cancer Research Center) for critical discussions. We thank EBI/EMBO's courses for catalyzing cooperation among coauthors.

**Funding** Microsoft Research and FAPESP grant 2009/53161-6. FAPESP fellowships (2010/14926-4, 2012/05392-1, 2010/15417-6). We thank the Programa de Pos-graduação em Genética FMRP-USP for CAPES travel grants and initial funding awarded to RRS.